# Atomic-like behaviors and orbital-related Tomonaga-Luttinger liquid in carbon nano-peapod quantum dots

J. Haruyama[1,5], J. Mizubayashi[1,5], I. Takesue[1,5], T. Okazaki[2,5], H. Shinohara[3,5], Y. Harada[4,5], Y. Awano[4,5]

[1] Aoyama Gakuin University, 5-10-1 Fuchinobe, Sagamihara, Kanagawa 229-8558 Japan
[2] National Institute of Advanced Industrial Science and Technology, Tsukuba, 305-8565, Japan
[3] Nagoya University, Furo-cho, Chigusa, Nagoya 464-8602 Japan
[4] Fujitsu Laboratory, 10-1 Wakamiya, Morinosato, Atsugi, Kanagawa 243-0197 Japan,
[5] JST-CREST, 4-1-8 Hon-machi, Kawaguchi, Saitama 332-0012 Japan

## Abstract

We report influence of encapsulated $C_{60}$ molecules on electron transport in carbon-nanotube peapod quantum dots. We find atomic-like behaviors with doubly degenerate electronic levels, which exist only around ground states, by single electron spectroscopy measured at low back-gate voltages ($V_{bg}$). Correlation with presence of nearly free electrons (NFEs) unique to the peapods is discussed. Moreover, we find anomalously high values of power α observed in power laws in conductance versus energy relationships, which are strongly associated with the doubly degenerate levels. It is revealed that the powers originate from Tomonaga-Luttinger liquid via the occupied doubly degenerate levels. Our observations clarify that the encapsulated $C_{60}$ molecules form doubly degenerate levels only at ground state in peapod quantum dots and do not eliminate a ballistic charge transport.

## 1. INTRODUCTION

Carbon nano-peapods, which are single-walled carbon nanotubes (SWNTs) encapsulating a series of fullerenes, such as $C_{60}$, $C_{70}$, and Gd@$C_{82}$ ($C_{82}$ encapsulating Gd) in their inner space (*1*)(*2*), have recently attracted considerable attention; this is because their remarkable nanostructures yield exotic electron states, charge transport, and one-dimensional (1D) quantum phenomena.

In $C_{60}$@(n,n) peapods, which are arm-chair type SWNTs encapsulating $C_{60}$ molecules, it has been predicted that electrons that were transferred from the SWNT accumulated in the space between the $C_{60}$ molecules and SWNTs, resulting in the so-called nearly free electrons (NFEs) (*3*). Hybridization of these NFEs with the π and σ orbitals of $C_{60}$ introduced four asymmetric subbands including the approximately doubly degenerate ground states in the $C_{60}$@(10,10) peapod as compared with the two subbands in conventional SWNTs (*3, 4*).

Measurements of semiconductive peapods encapsulating a series of Gd@$C_{82}$ by a scanning tunnel microscope revealed that a conduction band was periodically modulated around Gd@$C_{82}$ in a real space due to the hybridization of orbitals between the SWNT and Gd@$C_{82}$ (*5*). Moreover, electrical measurements of peapods encapsulating $C_{60}$ and Gd@$C_{82}$ indicated the possibility of the presence of variable range hopping (*6*). Refs. (*3*)–(*6*) at least suggested the presence of charge transfer and orbital hybridization between the encapsulated fullerenes and SWNTs.

On the other hand, SWNTs are within a 1D ballistic charge transport regime and have exhibited a variety of quantum effects, such as quantized energy levels, Tomonaga-Luttinger liquid (TLL) (*7 - 10*), and atomic-like behaviors as quantum dots (*11 - 14*). In particular, the behavior of TLL, which is a collective phenomenon arising from electron-electron interaction in 1D conductors, has been identified by observing power laws in relationships of conductance vs. energy in carbon nanotubes (CNs) (*7 - 10*). The reported correlation exponent *g*, which denotes the strength of an electron-electron interaction, was as low as ~ 0.2 and implied the presence of a strong repulsive Coulomb interaction in CNs. When CNs act as quantum dots, electron can be placed one by one. This single charging effect has shown even-odd effect, shell-filling to two spin-degenerate electronic states, and Kondo effect in CN quantum dots. How such phenomena are affected by encapsulating a series of fullerenes, however, has not yet been investigated in any carbon peapods.

## 2. EXPERIMENTAL RESULTS AND DUSCUSSIONS

For the present study, peapod field-effect transistors (FETs) were fabricated. An SEM top view revealed that the FETs included two bundles of peapods (*2*) as the channels. The number of peapods included in one bundle was estimated to be approximately 20 from measurements by the SEM, AFM, and TEM. Since the observed differential conductance was largely independent of the change in back gate voltage ($V_{bg}$), metallic transport in the present peapod was suggested.

First, the measurement result by single electron spectroscopy is shown in Fig.1(a) and (d). Figure 1(a) shows Coulomb diamonds that are the results in the low $V_{bg}$ region around $V_{bg}$ = 0V. The four diamonds, a sequence of one large diamond (shown as n = 4) followed by three smaller ones (shown as n = 1 – 3), are observable in Fig.1(a).

This sequence indicates possible presence of atomic-like behaviors with the doubly degenerate electronic levels only at ground states. However, the presence of only one set of this sequence of diamonds observable only around $V_{bg}$ = 0V is very different from the four fold diamonds periodically observed in SWNT (*13*) and multi-walled CN (MWNT) quantum dots (*14*). In conventional SWNT quantum dots, such unique degenerate levels are impossible, because each level is formed only from quantization of two subbands existing in bulk of a SWNT. In some cases, all levels are doubly degenerate like the four fold diamonds as observed in refs. [*13*] and [*14*]. In contrast, ref. [3] predicted presence of the four subbands with the doubly degenerate subband structures around ground state in the bulk of $C_{60}$@(10,10) peapod, although the influence of applying $V_{bg}$ and the dot structure were not taken into consideration in the theory. This degeneration might have to be also resolved in peapod quantum dot structure as well as the case of SWNT quantum dots, resulting in either all discrete electronic levels or all doubly degenerate levels [13], [14]. However, Fig.1(a) shows

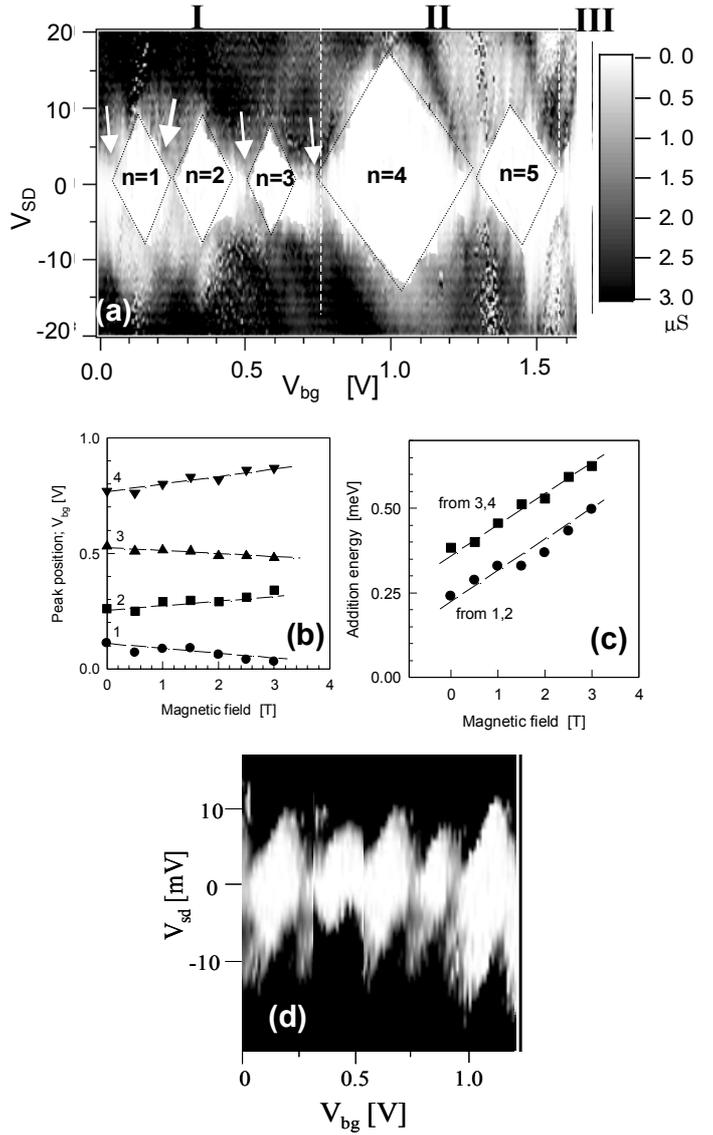

**Fig.1:** **(a)** Coulomb diamonds (white regions surrounded by the dotted lines) that show a sequence of one large diamond (n = 4) followed by three smaller ones (n = 1–3) of approximately equal size, observed in a peapod quantum dot at T = 1.5 K. The z-axis is the differential conductance with the magnitudes of which are indicated on the right side. *n* indicates the number of electrons confined in each diamond. Regions I, II, and III well correspond to those in Fig.4(b). **(b)** $V_{bg}$ shift of conductance peak positions (at $V_{bg}$ = 0.11 V, 0.26 V, 0.53 V, and 0.77 V around $V_{sd}$ = 0V shown by arrows in (a)) in Fig.1(a) as a function of magnetic field, B. **(c)** Addition energy obtained from each peak pair in (b) versus magnetic field, B. **(d)** Coulomb diamonds observed in an empty SWNT quantum dot (i.e., without encapsulating $C_{60}$ molecules).

neither of them and, then, only ground levels are doubly degenerate.

In order to confirm presence of doubly degenerate levels for Fig.1(a), we have investigated the back-gate-voltage shift of the linear-response conductance peaks (i.e., shown by arrows in Fig.1(a)) as a function of magnetic field, B, perpendicular to the tube axis. The result, Fig.1(b), reveals that adjacent peaks shift in opposite directions. This is a behavior of spin singlet state whose spins alternate as S=0 →1/2 →0 ……. and exist on the same orbital state, unlike a spin triplet state formed by Hund's rule. In Fig.1(c), we plot addition energy, which was deduced from the separation of adjacent peaks involving electrons on the same orbital in Fig.1(b) (i.e., peaks 1 and 2, peaks 3 and 4), as a function of magnetic field, B. A dashed line shows the result of the best fit of the data to $U_c + g_L \mu_B B$, where $\mu_B$ is the Bohr magneton and $g_L$ is the Lande factor, and gives $g_L$ = 1.96. This value of $g_L$ is approximately consistent with those mentioned in ref.[14]. Therefore, we conclude that Fig.1(a) implies presence of doubly degenerate electronic levels only at ground states and presence of atomic-like behaviors.

We interpret that these doubly degenerate levels originate from NFEs, which exists even in peapod quantum dots, because 1. Our empty SWNTs quantum dots, in which $C_{60}$s are not encapsulated and which have the same structures as those used for the present peapods, exhibited just even-odd effects (i.e., no-degenerate discrete levels) with $\Delta E$ = ~3 m eV, which is consistent with the previous studies in SWNT quantum dots (e.g., a $\Delta E$ of ~5 meV for a tube length of 100 ~ 200 nm) and the relationship $\Delta E = hv_F/2L$ (L and $v_F$ are the tube length and Fermi velocity, respectively), as shown in Fig.1(d) and 2. Even-odd effect similar to Fig.1(d) was detected at higher $V_{bg}$'s, which are continuous to the $V_{bg}$ region in Fig.1(a), in this peapod quantum dot. Hence, encapsulated $C_{60}$s yield the result of Fig.1(a). Ref.[3] predicted that the doubly degenerate subbands observable only at the ground state originated from the hybridization of orbitals in $C_{60}$ molecules and NFEs in bulk of peapods as mentioned in introduction and, hence, was a feature very unique to peapods. Based on analogy to these doubly degenerate subbands and compared with refs.[13][14], our results strongly indicate that the doubly degenerate electronic levels in Fig.1(a) also originate from NFEs and, hence, are very unique to peapod quantum dots. Disappearance of such degenerate levels in higher $V_{bg}$ region might be attribute to depletion of NFEs.

In addition, no atomic-like behavior was detected in the $-V_{bg}$ region. This result also supports correlation with NFEs, because this can be attributed to the modulation of the NFEs caused by increased electron density in the SWNT by applying $-V_{bg}$. This increase of electron density is caused by the excess electrons injected from Au electrodes as a result of a decrease in the potential difference between the peapod and Au electrode by the application of $-V_{bg}$. In contrast, applying $+V_{bg}$ increases this difference, thereby conserving electron density of the bulk of the peapod, resulting in the conserved NFEs. In this case, the tunneling current through the barrier dominates the carrier transport, thus exhibiting the above atomic-like behaviors.

Hence, peapod quantum dots can have a ballistic charge transport, only when a small $+V_{bg}$ was applied, in spite of presence of the encapsulated $C_{60}$s. Moreover, this result strongly indicates that the applied low $+V_{bg}$ play a major role in the modulation of neither NFE nor $C_{60}$, and that the low $+V_{bg}$ modulates only the position of the Fermi level in the SWNT for single electron injection. The significantly low value of the applied $V_{bg}$ enabled us to observe the electronic levels near the ground states, because of this no depletion of NFEs.

Here, the value of $U_c$ (the single-electron charging energy of the system) is approximately 6 ~ 10 meV in the small diamonds in Fig.1(a). This value for the present peapod with a length of 500 nm is approximately consistent with expectations based on a previous study of SWNT bundles (e.g. a $U_c$ ~25 meV for a tube length of 100–200 nm) (*11*). This result emphasizes that the encapsulated $C_{60}$ molecules do not contribute to the effective electrostatic capacitance for the single charging effect in this low $+V_{bg}$ region. Due to the high value of the applied $V_{bg}$, the outermost shell was depleted and the next outermost shell acted as a quantum dot in the MWNT (*14*). In contrast, this result implies that the SWNT is not fully depleted in this low $+V_{bg}$ region. This is consistent with the discussion mentioned above.

On the other hand, $\Delta E$ (the energy spacing in the dot), which was estimated from the difference in the size of the small and large diamonds shown in Fig.1(a), is ~13 meV. This value is the energy spacing between the doubly degenerate electronic-states and the upper state. However, it is much greater than expectations from the previous studies in SWNTs and the relationship $\Delta E = h v_F / 2L$ mentioned above. This may be also associated with NFEs.

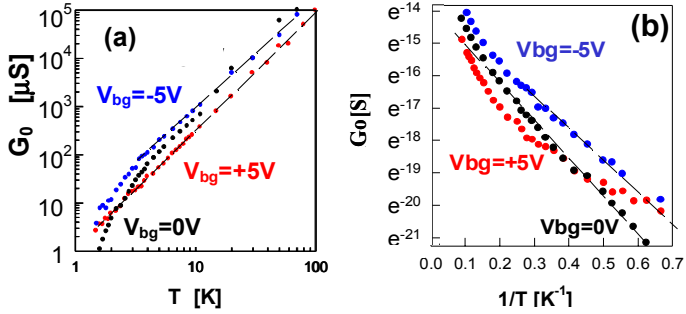

**Fig.2:** Zero-bias conductance ($G_0$) as a function of temperature for different back gate voltages ($V_{bg}$) on a doubly logarithmic scale **(a)** and an Arrhenius plot **(b)**. The solid lines are the results of data fitting.

Next, behaviors of TLL states are shown. Figure 2 (a) and (b) shows the relationship of zero-bias conductance ($G_0$) to the temperature for different values of $V_{bg}$ on a doubly logarithmic scale and an Arrhenius plot, respectively. $G_0$ monotonically decreases as temperature decreases for any values of $V_{bg}$. A distinct linear relationship (i.e., power law) with the power $\alpha$ as high as ~2.2 is observable for the entire temperature region only at $V_{bg}=+5V$ in Fig.2(a), whereas a linear relationship (i.e., thermal-activation type) is observable only at $V_{bg} = 0$ V in Fig.2(b), which indicates the presence of energy barriers at the interface of Au electrodes and peapods (*20*). At $V_{bg} = -5$ V, both these linear behaviors occur simultaneously with a boundary temperature at ~3 K. This polarity on $V_{bg}$ indicates the presence of asymmetric potential barriers between the source and drain sides of the Au/peapod interface, thus resulting in peapod quantum dots.

Figure 3 shows the relationships of differential conductance ($dI_{sd}/dV_{sd}$) to source-drain voltage ($V_{sd}$) on doubly logarithmic scales for one negative and three positive regions of $V_{bg}$. In the $-V_{bg}$ region, any dI/dV did not follow a linear relationship (a). On the contrary, saliently linear relationships with different $\alpha$ values are observable in the $+V_{bg}$ region. The behaviors are classified into three regions (b) - (d) as mentioned in the figure captions, showing an anomalously large $\alpha$ ($1.6 < \alpha < 12$).

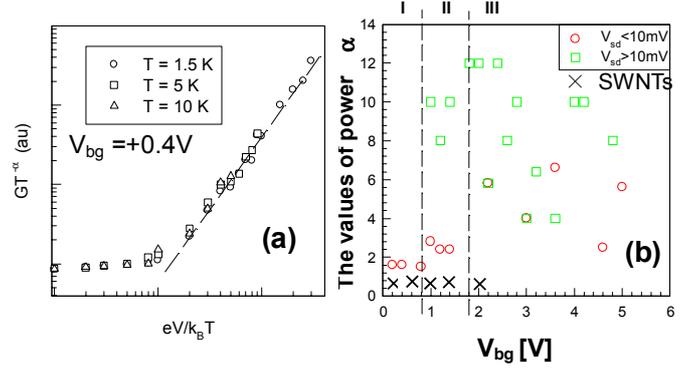

**Fig.3:** Relationships of differential conductance ($dI_{sd}/dV_{sd}$) to source-drain voltage ($V_{sd} \gg T = 1.5$ K measured) on doubly logarithmic scales for four different $V_{bg}$ regions. These power laws primarily appear just the nearest outside regions of diamonds (i.e., in the gray areas as shown by arrow) of Fig.1(a). Only the power law in the low $V_{sd}$ region at $V_{bg} = 1$ V appears in the n = 4 diamond. The liner lines are the results of data fitting. **(a):** For $-V_{bg}$ region. No dI/dV follows a linear relationship. **(b)-(d):** For $+V_{bg}$ regions. **(b):** The linearities with $1.6 < \alpha < 2$ are observable only at $V_{sd} < 0.01$ V. **(c):** Two linear relationships with different $\alpha$ ranges (i.e., $\alpha = 2 \sim 3$ and $\alpha = 8 \sim 10$ for $V_{sd} < 0.01$ V and $V_{sd} > 0.02$ V, respectively) are observable. **(d):** The linearities with $\alpha = 10 \sim 12$ are observable only at $V_{sd} > 0.01$ V.

Figure 4(a) shows the double-logarithmic plot of differential conductance divided by $T^\alpha$ as a function of $eV/k_B T$ measured at $V_{bg} = +0.4$V (included in Fig.3(b)) for three different temperatures. All data collapse on a single universal value with showing a saturation at $eV/k_B T < h v_F / L$, similar to that observed in ref.[8], but showing a large $\alpha$ similar to that in Fig.3(b). The values of $\alpha$ observed in all the $V_{bg}$ values including Fig.3 are shown in Fig.4(b). The differences in $\alpha$ among the three regions are apparent in this figure. Moreover, the values of $\alpha$ observed in empty SWNTs used for Fig.1(b) are also shown in Fig.4(b). All the values are less than 1, which is consistent with previous reports in SWNTs. Hence, Fig.4(a) is the results unique to peapods. In Fig.4(b), an $\alpha = \sim 2.4$, which is approximately consistent with that shown in Fig.2(a), is observable when $V_{sd} < 10$ mV and when $V_{bg} = 4.8$ V.

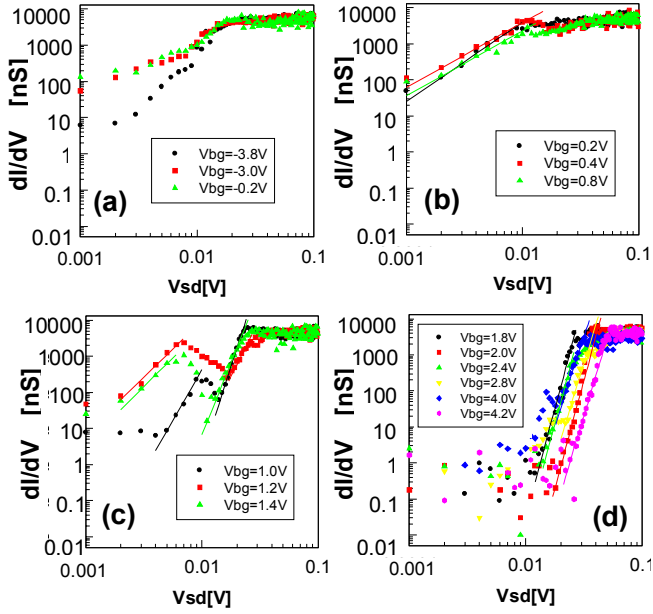

**Fig.4: (a)** $GT^{-\alpha}$ versus $eV/k_BT$ measures for measured for $V_{bg}=+0.4V$ of Fig.3(b). **(b)** Dependence of power $\alpha$ on different $V_{bg}$ values, estimated from Fig.3, in peapod and empty SWNT quantum dots. Three $V_{bg}$ regions (**(I):** $V_{bg} < 0.8$ V, **(II):** $0.8$ V $< V_{bg} < 1.8$ V, and **(III):** $1.8$ V $< V_{bg}$ corresponding to Figs.3(b), (c), and (d), respectively) clearly exist. Several values of $\alpha$ were added only in region III in addition to Fig.3(d).

The presence of power laws has been discussed as evidence for TLLs in CNs (*7 - 10*), as mentioned in the introduction. The values of $\alpha$ were very sensitive to the boundary conditions between the metal electrodes and CNs (*8*), namely the tunneling density of state; such as $\alpha^{bulk}= \sim 0.3$ and $\alpha^{end} =2\ \alpha^{bulk}$ for the tunneling from an Au electrode to the bulk and to the end of CNs within the large-channel number TLL states, respectively (*8*). The formulas of $\alpha$ for each tunneling were also given by $\alpha^{bulk} = (g^{-1} + g - 2)/8$ and $\alpha^{end} = (g^{-1} - 1)/4$. However, it should be noted that even the maximum value of $\alpha$ reported in CNs to date is approximately 1.25, except for refs.(*9,16*). Therefore, we imply that the $\alpha$ values of 1.6 ~ 12 observed in Figs.2 – 4 are anomalously large in comparison with the $\alpha$ values reported thus far in conventional TLLs (*9*). The junction structures in this study, in which the ends of the peapod bundles were placed under an Au electrode, should have shown a maximum $\alpha^{end}$ of only ~0.6. In fact, the empty SWNTs have exhibited $\alpha = \sim 0.8$ even at the maximum case as explained for Fig.4(b) above.

Here, it should be noted that the power laws shown in Fig.3 exist at each $V_{bg}$ in the gray areas, which are just the nearest outside regions of the Coulomb diamonds shown in Fig.1(a). Hence, in order to understand the power laws with anomalously large $\alpha$, it is crucial to note that the three positive $V_{bg}$ regions shown in Fig. 3(b) - (d) and Fig.4(b) are in good agreement with the three $V_{bg}$ regions classified by using the boundaries of diamonds as shown in Fig.1(a). These results clearly indicate that the power law behaviors with the large values of $\alpha$ are strongly associated with the number of (partially) occupied electronic levels, $N$, in each diamond

The correlation of power laws and $\alpha$ (TLLs) with the electronic-state filling effect (i.e., orbital filling effect) in CNs has not yet been reported in previous studies. Only a single study (*17*), however, predicted that a small $g$ and large $\alpha$ could be obtained from the large $N$ in peapods. The theory predicted $g = (1+2Nv_q/\pi \hbar v_F)^{-1/2}$ for armchair CNs, where $N$ and $v_q$ are the number of (partially) occupied symmetric subbands with degenerate Fermi vector waves and the same band width, and the electron-electron interaction matrix element, respectively. If the subbands are asymmetric and each of them crosses the Fermi level only once, $N$ can be replaced by $N/2$. This holds true for the subbands of the $C_{60}@(10,10)$ peapod in this study.

We quantitatively examine the validity of this theory for the present measurement by replacing $N$ to the number of electronic states and using the same value of $v_q$. The value of $g = 0.135$ is obtained from $\alpha^{end} = (g^{-1} – 1)/4$ (*8*) using $\alpha = 1.6$ that is observed in region I (N=2). The value of $v_q$ can be estimated by substituting these values of $g$ and N in $g = (1 + 2(N/2)v_q/\pi \hbar v_F)^{-1/2}$ (*17*). Then, $g = 0.11$ and $0.099$ are respectively obtained for N = 3 and N = 4 by substituting the estimated value of $v_q$ in $g = (1 + 2(N/2)v_q/\pi \hbar v_F)^{-1/2}$. The value of $g = 0.11$ for N = 3 is approximately in good agreement with $g = 0.082$ estimated from $\alpha^{end} = (g^{-1} - 1)/4$ by using $\alpha = 2.8$ that is observed in the portion of region II with low $V_{sd}$ values.

On the other hand, this $g$ value is irrelevant to $\alpha = 8 \sim 10$ that is observed in the portion of region II with high $V_{sd}$ values. Moreover, the $g$ value of 0.099 leads to $\alpha = 2.28$ for N = 4, which is significantly less than the values of $\alpha > 10$ observed in the region III at high $V_{sd}$ values. These indicate that different values of $v_q$ should be used for the case of higher $V_{sd}$. Because strength of electron-electron interaction varies from low to high $V_{sd}$s, this is reasonable. When different values of $v_q$ are used for large $N$, $\alpha = 10$ (for N = 3) and 12 (for N = 4) could be obtained from the values of $2v_q/\pi \hbar v_F = 1160$ and 1250, respectively, $g = (1 + 2(N/2)v_q/\pi \hbar v_F)^{-1/2}$, and $\alpha^{end} = (g^{-1} - 1)/4$.

Consequently, the theory (17) is quantitatively relevant when N = 2 and 3 (at lower $V_{sd}$) under the same value of $v_q$ and N = 3 (at higher $V_{sd}$) and 4 under the larger values of $V_q$. This indicates that the presence of two power laws observed in Fig.3(c) is attributed to change in $v_q$ due to increase in $V_{sd}$. Therefore, we conclude that the power laws with large values of $\alpha$ (1.6 < $\alpha$ < 12) can be attributed to the TLL via the occupied doubly degenerated electronic levels, which are located near the ground states unique to the peapod quantum dots. Here, presence of TLL states means also presence of a ballistic charge transport. Because power laws were not detected in $-V_{bg}$ region, this is consistent with absence of the atomic-like behaviors in $-V_{bg}$ region. Only when low $+V_{bg}$ is applied and NFEs are not modulated, carbon nano-peapods can show a ballistic charge transport. Further investigation is, however, required to reconfirm the values of $v_q$. As shown in Fig.3(c)(d), the $V_{sd}$ regions exhibiting power laws are very small in $V_{sd}$'s > 10 mV, i.e., at best half decade and the high values of power do not become constant in the region III. Hence, other interpretation will be possible.

The observations reported in this study strongly suggest that the 1D quantum phenomena observable in peapods are very exotic and associated with the electronic states (formed via NFEs) unique to the peapods. Further investigation is required in order to develop a comprehensive understanding of these phenomena.

### 3.Acknowledgments.
We acknowledge T.Nakanishi, S.Tarucha, W.Izumida, and M.Thorwart for valuable discussions